\documentstyle [12pt]{article}
\begin{document}


\newcommand{\Eqn}{\begin{equation}}
\newcommand{\eqn}{\end{equation}}
\newcommand{\EqnA}{\begin{eqnarray}}
\newcommand{\eqnA}{\end{eqnarray}}
\newcommand{\Eq}[1]{\mbox{Eq.\,(\ref{eqn:#1})}}
\newcommand{\EqRef}[1]{(\ref{eqn:#1})}
\newcommand{\Eqs}[1]{\mbox{Eqs.\,(\ref{eqn:#1})}}
\newcommand{\EqsRef}[2]{(\ref{eqn:#1}\,{#2})}

\newcommand{\MinusA}[1]{\nonumber \\ & & \mbox{} \!\!\!\!\!\!\! -}
\newcommand{\PlusA}[1]{\nonumber \\ & & \mbox{} \!\!\!\!\!\!\! +}
\newcommand{\EquA}{&\!\!\!\!=\!\!\!\!&}
\newcommand{\ApprA}{&\!\!\!\!\approx\!\!\!\!&}
\newcommand{\EquAC}{\nonumber \\ \EquA}
\newcommand{\EquACn}{\\ \EquA}
\newcommand{\Minus}{\MinusA{0em}}
\newcommand{\Plus}{\PlusA{0em}}
\newcommand{\minus}{\!-\!}
\newcommand{\plus}{\!+\!}

\newcommand{\Prop}{\sim}

\newcommand{\figlist}[1]{
   \newcounter{fig}
   \begin{list}{\bf Figure \arabic{fig}:}{\usecounter{fig}
      \labelwidth2.0cm \leftmargin2.5cm \labelsep0.5cm \rightmargin1cm
      \parsep0.5ex plus0.2ex minus0.1ex \itemsep0ex plus0.2ex \rm}
      #1
   \end{list}
   }

\newcommand{\tablist}[1]{
   \newcounter{tab}
   \begin{list}{\bf Table \arabic{fig}:}{\usecounter{fig}
      \labelwidth2.0cm \leftmargin2.5cm \labelsep0.5cm \rightmargin1cm
      \parsep0.5ex plus0.2ex minus0.1ex \itemsep0ex plus0.2ex \rm}
      #1
   \end{list}
   }


\title
{\LARGE\bf
                Resummation of classical and semiclassical
                          periodic orbit formulas
}
\author {
                    Bruno Eckhardt and Gunnar Russberg
\\ \ \\
\it               Fachbereich Physik der Philipps-Universit\"{a}t,\\
\it                         Renthof 6, D-3550 Marburg
}

\maketitle

\abstract
{
The convergence properties of cycle expanded periodic orbit expressions
for the spectra of classical and semiclassical time evolution operators
have been studied for the open three disk billiard. We present evidence
that both the classical and the semiclassical Selberg zeta function
have poles. Applying a Pad\'{e} approximation on the expansions of the
full Euler products, as well as on the individual dynamical zeta
functions in the products, we calculate the leading poles and the zeros
of the improved expansions with the first few poles removed.
The removal of poles tends to change the simple linear exponential
convergence of the Selberg zeta functions to an $\exp\{-n^{3/2}\}$
decay in the classical case and to an $\exp\{-n^2\}$
decay in the semiclassical case.

The leading poles of the $j$th dynamical zeta function are found to equal
the leading zeros of the $j+1$th one: However,
in contrast to the zeros, which
are all simple, the poles seem without exception to be {\em double}\/.
The poles are therefore in general {\em not}\/ completely
cancelled by zeros,
which has earlier been suggested. The only
complete cancellations occur in
the classical Selberg zeta function between
the poles (double) of the first
and the zeros (squared) of the second dynamical zeta function.

Furthermore, we find strong indications that poles are responsible for
the presence of spurious zeros in periodic orbit quantized spectra
and that these spectra can be greatly improved
by removing the leading poles, e.g.\ by using the Pad\'{e} technique.
}

\bigskip\bigskip


\newpage

\section{Introduction}

Trace formulas in chaotic dynamical systems relate phase space averages
to sums over periodic orbits \cite{Ruelle,RE,AACI,AACII,CEcl,Varenna}.
Exponentiated trace formulas give rise to Selberg type
zeta functions, named after corresponding expressions arising in studies
of billiards on surfaces of negative curvature \cite{Selberg,BV}.
Selberg zeta functions
factorize further into products of dynamical zeta functions, each one
being an infinite product over all primitive nonrepeated periodic orbits
(ppo) of the system. Finally, the term cycle expansion refers to a
certain expansion and truncation
of (dynamical and Selberg) zeta functions into polynomials. Whereas the
original trace formulas and the infinite products
have the same convergence and analyticity properties, cycle expanded
periodic orbit expressions typically
converge much better \cite{AACI,AACII}.

Calculations can be improved, if the pole structure is known \cite{CCR}.
Typically, dynamical zeta functions will have poles; for Selberg
zeta functions, one can advance arguments \cite{AACI,Aurell}
that they should be entire
and thus ideally suited for numerical purposes.
We here present quantitative results on the analyticity properties
of zeta functions for a 2d conservative
dynamical system, a point particle elastically scattered off three disks
placed symmetrically in the plane \cite{Eck87,gaspard,CE89,Pinball}.
This system is ideally suited for such an
investigation since (for sufficiently
separated disks) it is a hyperbolic
system with a good symbolic coding
(complete binary, once the symmetries
are factored out). Periodic orbits can conveniently and accurately be
computed.

We proceed with a formal definition of the objects investigated:
Let $p$ label all primitive non repeated periodic orbits, $n_{p}$ their
symbolic length, $\mu_{p}$ the Maslov
index (in a billiard $\mu_{p}=2n_{p}$),
and $J_{p}$ the linearization perpendicular to the orbit with
$\Lambda_{p}$ the eigenvalue of largest
absolute value. For the two degree
of freedom system considered here $J_{p}$ is a $2\times2$ matrix of
determinant 1 so that the other eigenvalue is $1/\Lambda_{p}$.
We then consider the classical Selberg zeta function \cite{CEcl}
\Eqn
         Z(z)
         =\exp\left\{-\sum_{p}
                      \sum_{r=1}^{\infty}\frac{z^{r n_{p}}}{r}
                      \frac{1}{|\det(1-J_{p}^{r})|}\right\}
         =\prod_{j=0}^{\infty}\left[
          \zeta_{j}^{-1}(z)\right]^{j+1}
\label{eqn:Zcl}
\eqn
with
\Eqn
         \zeta_{j}^{-1}(z)
         =\prod_{p}
              \left(1-z^{n_{p}}|\Lambda_{p}|^{-1}
              \Lambda_{p}^{-j}\right)
         \ \ \ ,   \label{eqn:zcl}
\eqn
and the semiclassical Selberg zeta function \cite{Gut0,Gutz,Vor}
\Eqn
         Z(z)
         =\exp\left\{-\sum_{p}
             \sum_{r=1}^{\infty}\frac{z^{r n_{p}}}{r}
\frac{e^{-ir\mu_{p}\pi/2}}{|\det(1-J_{p}^{r})|^{1/2}}\right\}
         =\prod_{j=0}^{\infty}
             \zeta_{j}^{-1}(z)
                   \label{eqn:Zqm}
\eqn
with
\Eqn
         \zeta_{j}^{-1}(z)
         =\prod_{p}
             \left(1-z^{n_{p}}e^{-i\mu_{p}\pi/2}|\Lambda_{p}|^{-1/2}
             \Lambda_{p}^{-j}\right)
         \ \ \ ,   \label{eqn:zqm}
\eqn
all as functions of $z$.
We avoid use of additional labels distinguishing classical and
semiclassical zeta functions and hope that it is clear from the
context which one is meant.
Straightforward formal manipulations allow to express
each function (1)-(4) as a power series $\sum C_{n}z^{n}$ in $z$,
which, when truncated, yields the cycle expansion.
Where needed, we will abbreviate with $t_p$ the contributions of
periodic orbits to the dynamical zeta function with $j=0$ so that
$\zeta^{-1}_j = \prod_p (1-t_p\Lambda_{p}^{-j})$.

The above expressions are correct for maps, the periods of orbits
being integers. For flows one would replace
$z^{n_p}$ by $z^{n_p}\exp\{i\omega T_p\}$ in the classical case or
by $z^{n_{p}}\exp\{iS_{p}(E)/\hbar\}$
in the semiclassical case, expand
in a power series in $z$ and consider
the final result as a function of
frequency $\omega$ or energy $E$, respectively, for $z=1$.
In billiards, the action is given by $S_{p}(E)/\hbar=L_{p}k(E)$,
where $L_{p}$ is the geometrical length of the ppo
and $k(E)=\sqrt{2mE}/\hbar$ is the wave number.
In addition to Maslov phases there can be further phases due to
symmetries \cite{Rob,Laur,CESym}.
For the case of the three disks, the above expressions
are correct in the $A_1$-representation;
in the $A_2$-representation there is
an additional phase $i\pi$ if $n_p$ is odd.

Periodic orbits for the three disk system have been
computed using Newton's
method on two different maps, one based on direct description (impact
parameter, scattering cycle) of collisions with the disks and one
based on stationarity of action.
The computations were done for several values
of the ratio $\rho\equiv d/R$,
where $d$ is the disk separation and $R$ the
disk radius. Symmetry reduced orbits up to symbolic
length 13 have been found in double
precision numerics with relative accuracy
$10^{-14}$. The exponential of \EqRef{Zcl} and \EqRef{Zqm} was computed
using all orbits and repetitions satisfying $n_p r \le N$ and
then expanded in a series $\sum C_{n}z^{n}$
using the recurrence relations
of Plemelj and Smithies\cite{Pl,Sm,RS}.

Fig.~1 shows the results for the classical Selberg
zeta function. An apparently
faster than exponential decay is observed.
In contrast, the semiclassical
Selberg zeta function seems to decay faster
than exponential for the first
three or four terms, but then settles for an
exponential decay, see Fig.~2.
In the following we will explain this difference
in behaviour,
show its effects in calculations and demonstrate how this knowledge
can be used to improve calculations.

We begin in section~2 with a detailed discussion of the convergence
of cycle expanded zeta functions, including numerical
results for the three disk system. In section~3 we turn to methods
for identification and removal of poles. The effect of poles
when calculating quantum resonances is discussed in section~4.
We conclude with a short summary and
comments on other systems in section~5.

\section{Convergence estimates}

Some insight into the behaviour of the cycle expansions of
\EqRef{Zcl} and \EqRef{Zqm} may be obtained by considering the
special case of a complete binary code with
$\Lambda_{0}\Prop\Lambda_{1}\Prop\Lambda$ and a factorization of the
eigenvalues of the longer periodic orbits
$\Lambda_{p}\Prop\Lambda_{0}^{n_{0}}\Lambda_{1}^{n_{1}}$, where $n_{0}$ and
$n_{1}$ are the numbers of zeros and ones in the symbolic description of $p$.
Then the classical dynamical zeta functions take on the form
\Eqn
      \zeta_{j}^{-1}(z)\Prop
             \left(1-2z|\Lambda|^{-1}\Lambda^{-j}\right)
         \ \ \ .                                               \label{eqn:zcl2}
\eqn
Expanding the product on $j$ in a power series in $z$, one finds
\Eqn
      Z(z)=
      \prod_{j}\left[\zeta_{j}^{-1}(z)
                 \right]^{j+1}\Prop\sum d_{n}z^{n}
\eqn
with $d_{n}\Prop\Lambda^{-n^{3/2}}$ (see below). Similarly, in the
semiclassical case one finds
\Eqn
      \zeta_{j}^{-1}(z)\Prop
             \left(1-2z|\Lambda|^{-1/2}\Lambda^{-j}\right)
                                                               \label{eqn:zqm2}
\eqn
and
\Eqn
      Z(z)=
      \prod_{j}\zeta_{j}^{-1}(z)
      \Prop\sum d'_{n}z^{n}
\eqn
with $d'_{n}\Prop\Lambda^{-n^{2}}$ (essentially due to the Euler product
formula, see e.g.~eq.~(89.18.3) in Ref.~\cite{Hansen}).
Because of this rapid decay, these functions are free of poles.

In the general hyperbolic case, the expansion of dynamical zeta functions
does not stop with the linear term but rather continues with exponentially
decaying coefficients, $c_{n}\Prop \beta\,^{n}$ with $|\beta|<1$.
Summing this geometrical series one finds a pole at
$z=\beta^{-1}$. In ref.~\cite{AACI,Aurell} $\beta$ has been related to the
Lyapunov exponents which
suggests that the poles of $\zeta_{j}^{-1}$ should be compensated by the
zeros of $\zeta_{j+1}^{-1}$.
To see what terms have to compensate in order to provide faster than
exponential decay of the coefficients in the Selberg type zeta
function, let us consider the classical and semiclassical cases
in some more detail.

\newcommand{\fj}[2]{f^{(#1)}_{#2}}
\newcommand{\cj}[2]{c^{(#1)}_{#2}}
\newcommand{\fjn}[2]{f^{(#1)}_{i_{#2}}}
\newcommand{\cjn}[2]{c^{(#1)}_{i_{#2}}}

\subsection{The semiclassical case}

With each dynamical zeta function expanded in a power series in $z$, the
semiclassical Selberg zeta function \EqRef{Zqm} looks like
\EqnA
    Z(z) \EquA \prod_{j}\zeta_{j}^{-1}(z)
                                                            \label{eqn:Zfc}
        \EquACn (1\minus\fj{0}{1}z\minus\cj{0}{6}z^{2}
             \minus\cj{0}{9}z^{3}\minus\ldots)
    (1\minus\fj{1}{3}z\minus\cj{1}{10}z^{2}\minus\ldots)
    (1\minus\fj{2}{5}z\minus\cj{2}{14}z^{2}\minus\ldots)\ldots
             \nonumber
\eqnA
As is often convenient for dynamical zeta functions we distinguish
curvature terms $c_{i_{n}}$ of orbits grouped together with shorter shadowing
orbits and fundamental contributions $f_{i}$ of the shortest orbits, which by
definition are not approximated by other orbits. Superscript labels here
indicate the order $j$ of the zeta function. Subscripts indicate the size of
the terms in powers of $|\Lambda|^{-1/2}$, where $|\Lambda|$ is a typical
instability of the shortest orbits (obviously, some uniformity in the
Lyapunov exponents is assumed here). For instance, $\fj{0}{1}$ comes
from terms of the form $|\Lambda_{p}|^{-1/2}e^{-i\mu_{p}\pi/2}$ in
$\zeta_{0}^{-1}$, whence its subscript equals 1. Fundamentals $\fj{j}{i}$
with higher $j$ have additional powers $|\Lambda_{p}|^{-j}$ and thus
$i=1+2j$. The order of the curvature corrections
$c_{i_n}^{(j)} z^n$ is determined
by two factors: the typical size of the terms contributing
(about $|\Lambda|^{-n/2-jn}$) and an additional factor $\Prop\Lambda^{-n}$
due to exponential shadowing of long orbits by short approximands, thus
$i_{n}=\left(3+2j\right)n$.

To estimate the convergence behaviour, we will now expand
\EqRef{Zfc} in $z$, $Z(z)=\sum C_{n}z^{n}$, and evaluate
the leading order contributions in each coefficient $C_{n}$.
One sees that there is a considerable difference in order of magnitude between
the fundamental term and the first curvature correction, so to begin with
one would expect to find significant contributions from fundamentals only.
In the ideal two scale approximation
$\Lambda_{p}=\Lambda_{0}^{n_{0}}\Lambda_{1}^{n_{1}}$ all curvatures are
identically zero, $\cjn{j}{n}=0$, and the expansion looks as follows
\EqnA
     Z(z) \EquA  \left(1\minus\fj{0}{1}z\right)
                 \left(1\minus\fj{1}{3}z\right)
                 \left(1\minus\fj{2}{5}z\right)\ldots
          \EquAC 1-\left(\fj{0}{1}\plus\fj{1}{3}
                   \plus\fj{2}{5}\plus\ldots\right)z
          \Plus  \left(\fj{0}{1}\fj{1}{3}\plus\fj{0}{1}\fj{2}{5}
                 \plus\fj{0}{1}\fj{3}{7}\plus\fj{1}{3}\fj{2}{5}
                 \plus\ldots\right)z^{2}
          \Minus \left(\fj{0}{1}\fj{1}{3}\fj{2}{5}
                  \plus\fj{0}{1}\fj{1}{3}\fj{3}{7}
                  \plus\ldots\right)z^{3}
                  +\ldots \equiv \sum_{n=0}^{\infty} d'_{n}z^{n}
          \ \ \ .
\eqnA
We notice that the leading terms grow in order ($I_{n}$) like $I_{1}=1$,
$I_{2}=1+3=4$, $I_{3}=1+3+5=9$, $\ldots$,
i.e., $\log d'_{n}\Prop I_{n}=\sum_{j=0}^{n-1}(1+2j)=n^{2}$, a quadratic
exponential convergence.

In the full evaluation of \EqRef{Zfc} we cannot expect the purely fundamental
product terms to be leading forever. Arranging products like above
acording to the
sum of the lower indices and calling the leading fundamental terms
$F_{n^{2}}$ we find the following expansion coefficients $C_{n}$:
\EqnA
     C_{0} \EquA 1
                                                  \nonumber\\ [0.8ex]
     C_{1} \EquA F_{1}
                 -\fj{1}{3}-\fj{2}{5}+O(7)
                                                  \nonumber\\ [0.8ex]
     C_{2} \EquA F_{4}
                 +\fj{0}{1}\fj{2}{5}-\cj{0}{6}
                 +\fj{0}{1}\fj{3}{7}+\fj{1}{3}\fj{2}{5}
                 -\cj{0}{9}
                 +O(10)
                                                  \nonumber\\ [0.8ex]
     C_{3} \EquA F_{9}
                 -\cj{0}{9}+\fj{1}{3}\cj{0}{6}
                 -\fj{0}{1}\fj{1}{3}\fj{3}{7}
                 +\fj{2}{5}\cj{0}{6}
                 +\fj{0}{1}\cj{1}{10}
                 +O(12)
                                                  \nonumber\\ [0.8ex]
     C_{4} \EquA -\cj{0}{12}+\fj{1}{3}\cj{0}{9}
                 +\fj{2}{5}\cj{0}{9}
                 -\fj{1}{3}\fj{2}{5}\cj{0}{6}
                 +F_{16}
                 +\fj{0}{1}\cj{1}{15}
                 +\ldots
                 +O(18)
                                                  \nonumber\\ [0.8ex]
     C_{5} \EquA -\cj{0}{15}+\fj{1}{3}\cj{0}{12}
                 +\fj{2}{5}\cj{0}{12}
                 -\fj{1}{3}\fj{2}{5}\cj{0}{9}
                 +\ldots
                 +F_{25}
                 +\ldots
                 +O(27)
\eqnA
etc, where $O(n)$ indicates terms of size $\sim \Lambda^{-n}$.
Up to and including $n=3$ the convergence is unconditionally quadratic as
in the ideal situation. At larger $n$, however, due to the simple
exponential decay of the pure curvatures, these as well as mixed curvature
and fundamental cross product terms have outgrown the pure
fundamental ones. Unless there now exist efficient additional cancellations
within complexes of the form
\Eqn
    \cj{0}{3n}-\fj{1}{3}\cj{0}{3n-3} \ \ ,
                                                      \label{eqn:canc_qm}
\eqn
raising their order to at least $n^{2}$,  a sudden change
in the convergence behaviour of the semiclassical Selberg zeta function around
$n=4$ is to be expected. This is indeed what is observed in Figure 2.

\subsection{The classical case}

Following the procedure in the semiclassical treatment above we now analyse the
slightly more complicated classical Selberg zeta function \EqRef{Zcl},
\EqnA
    Z(z) \EquA \prod_{j}\left[\zeta_{j}^{-1}(z)\right]^{j+1}
                                                            \label{eqn:Zclfc}
        \EquACn (1\minus\fj{0}{1}z\minus\cj{0}{4}z^{2}
             \minus\cj{0}{6}z^{3}\minus\ldots)
    (1\minus\fj{1}{2}z\minus\cj{1}{6}z^{2}\minus\ldots)^{2}
    (1\minus\fj{2}{3}z\minus\cj{2}{8}z^{2}\minus\ldots)^{3}\ldots
             \nonumber
\eqnA
Since in the classical case weights are proportional to powers of
$|\Lambda_{p}|^{-1}$, subscripts now indicate the size of the terms in
powers of $|\Lambda|^{-1}$ rather than $|\Lambda|^{-1/2}$.

The convergence behaviour in the ideal case $\cjn{j}{n}=0$ is found as follows:
>From a straightforward expansion of
\Eqn
     Z(z) = \left(1\minus\fj{0}{1}z\right)
                \left(1\minus\fj{1}{2}z\right)^{2}
                \left(1\minus\fj{2}{3}z\right)^{3}\ldots
           \equiv \sum_{n=0}^{\infty}d_{n}z^{n}
\eqn
one obtains the expansion coefficients
\EqnA
     d_{0} \EquA 1
                                                  \nonumber\\ [0.8ex]
     d_{1} \EquA -\fj{0}{1}
                 +\ldots=O(1)
                                                  \nonumber\\ [0.8ex]
     d_{2} \EquA 2\fj{0}{1}\fj{1}{2}
                 +\ldots=O(3)
                                                  \nonumber\\ [0.8ex]
     d_{3} \EquA -\fj{0}{1}\fj{1}{2}\fj{1}{2}
                 +\ldots=O(5)
                                                  \nonumber\\ [0.8ex]
     d_{4} \EquA 3\fj{0}{1}\fj{1}{2}\fj{1}{2}\fj{2}{3}
                 +\ldots=O(8)
                                                  \nonumber\\ [0.8ex]
     d_{5} \EquA -3\fj{0}{1}\fj{1}{2}\fj{1}{2}\fj{2}{3}
                 \fj{2}{3}
                 +\ldots=O(11)
                                                  \nonumber\\ [0.8ex]
     d_{6} \EquA \fj{0}{1}\fj{1}{2}\fj{1}{2}\fj{2}{3}
                 \fj{2}{3}\fj{2}{3}
                 +\ldots=O(14)
                                                  \nonumber\\ [0.8ex]
     d_{7} \EquA -4\fj{0}{1}\fj{1}{2}\fj{1}{2}\fj{2}{3}
                 \fj{2}{3}\fj{2}{3}\fj{3}{4}
                 +\ldots=O(18)
\eqnA
etc. The growth rule should be obvious: From the $j$th zeta function
(counting $\zeta_{0}^{-1}$ as the 1st), there are $j$ consecutive
contributions to the leading order terms, each increasing the order
of magnitude by an amount $j$, i.e., $j=j_{n}$ grows by one
over an interval of length $\Delta n=j_{n}$ and the total growth in
order $I_{n}$ is $\Delta I_{n}=I_{n+\Delta n}-I_{n}=j^{2}_{n}$.
For large $n$ one thus ends up with the following differential equations:
\EqnA
     \frac{dn}{dj} \EquA j                               \label{eqn:Diff1}
       \\ [0.8ex]
     \frac{dI}{dn} \EquA j                               \label{eqn:Diff2}
     \ \ \ .
\eqnA
\Eq{Diff1} gives $j_{n}=n^{1/2}$, which is inserted into \Eq{Diff2}.
The solution of the resulting equation is the sought for asymptotic relation
$I_n\Prop n^{3/2}$.

The full evaluation of \Eq{Zclfc} gives
\EqnA
     C_{0} \EquA 1
        \nonumber \\ [0.8ex]
     C_{1} \EquA F_{1}-2\fj{1}{2}-3\fj{2}{3}
        +O(4)
        \nonumber \\ [0.8ex]
     C_{2} \EquA F_{3}
        -\cj{0}{4}-\fj{1}{2}\fj{1}{2}+3\fj{0}{1}\fj{2}{3}
        +6\fj{1}{2}\fj{2}{3}+4\fj{0}{1}\fj{3}{4}
        +O(6)
        \nonumber \\ [0.8ex]
     C_{3} \EquA F_{5}
        -\cj{0}{6}+2\fj{1}{2}\cj{0}{4}-6\fj{0}{1}\fj{1}{2}\fj{2}{3}
        +2\fj{0}{1}\cj{1}{6}+3\fj{2}{3}\cj{0}{4}
        -\ldots
        +O(8)
        \nonumber \\ [0.8ex]
     C_{4} \EquA F_{8}
        -\cj{0}{8}+2\fj{1}{2}\cj{0}{6}-\fj{1}{2}\fj{1}{2}\cj{0}{4}
        -2\fj{0}{1}\fj{1}{2}\cj{1}{6}
        +3\fj{2}{3}\cj{0}{6}
        +\ldots
        +O(10)
        \nonumber \\ [0.8ex]
     C_{5} \EquA -\cj{0}{10}+2\fj{1}{2}\cj{0}{8}-\fj{1}{2}\fj{1}{2}\cj{0}{6}
        +F_{11}
        +\ldots
        +O(12)
\eqnA
etc.
The leading order terms in each $C_{n}$ which have
to compensate in order to get the faster than exponential convergence
$n^{3/2}$ are now of the form
\Eqn
     \cj{0}{2n}-2\fj{1}{2}\cj{0}{2n-2}+\fj{1}{2}\fj{1}{2}\cj{0}{2n-4} \ \ .
                                                      \label{eqn:canc_cl}
\eqn
Things work out nicely until order 4; beginning from order 5
additional cancellations are needed. As figure 1 shows, these seem
to occur in the classical case.

\subsection{Numerical estimate of curvatures}

A rather crude estimate of the {\em individual}\/ curvature terms,
$
   \cj{j}{p}\Prop t_{p}\Lambda_{p}^{-(j+1)}
$,
where
$
   t_{p}=|\Lambda_{p}|^{-1}
$
in the classical and
$
   t_{p}=|\Lambda_{p}|^{-1/2}
$
in the semiclassical case, was used above to obtain the correct order of
magnitude for the {\em full}\/ curvatures $\cjn{j}{n}$, each of which
being a sum over individuals with the same symbol length.
One would be able to benefit more from the results of the preceding
sections if there were a better estimate of the individual curvatures
$\cj{0}{p}$ in $\zeta^{-1}_{0}$. Consider again a system with binary
symbolic dynamics:

First note that due to uncertainty in the building of complexes like
$
   \cj{0}{00011}=t_{00011}-t_{0001}t_{1}-t_{0}t_{0011}+t_{0}t_{001}t_{1}
$
and
$
   \cj{0}{00101}=t_{00101}-t_{001}t_{01}
$
it appears necessary to collect curvatures with the same number of
zeros and ones into a single term
\Eqn
   c_{nm}\equiv\sum_{p}\delta_{n,n_{0}}\delta_{m,n_{1}}\cj{0}{p} \ \ ,
\eqn
where the Kronecker $\delta$'s select primitive periodic orbits with
number of symbols $n_{0}=n$ and $n_{1}=m$.
After this precaution one may make the following {\em ansatz}\/,
\Eqn
   c_{n_{0}n_{1}}\equiv\alpha_{n_{0}n_{1}}
            \left(t_{0}\Lambda_{0}^{-1}\right)^{n_{0}}
            \left(t_{1}\Lambda_{1}^{-1}\right)^{n_{1}} \ \ ,
\eqn
hoping that the essential stability dependence has been correctly
extracted, so that the prefactors $\alpha_{n_{0}n_{1}}$ depend only
weakly on stability.
We aim at finding an approximation for the prefactors $\alpha_{n_{0}n_{1}}$
better than $\alpha_{n_{0}n_{1}}\Prop 1$.

We have calculated the prefactors $\alpha_{n_{0}n_{1}}$
up to symbol length 6 for the open three disk system,
with values of $\rho$ ($=d/R$) ranging from 2.5 to 6.0, and with
weights given by $t_{p}=|\Lambda_{p}|^{-D}$. The results of the
calculations are presented in Table~1 ($D=1/2$, semiclassical case)
and Table~2 ($D=1$, classical case).
The values for $D=1$ are roughly twice those for $D=1/2$ and some
variation in parallel with the instabilities
$\Lambda_0$ and $\Lambda_1$ are noticable.
The dependence on $n_{0}$ and $n_{1}$
seems to be roughly binomial
with an additional factor $n_{0}+n_{1}$;
we conclude that the data in Tables 1 and 2 should be more or less well
approximated by the following formula:
\Eqn
   \alpha_{n_{0}n_{1}}\approx
       D h_{D}(\Lambda_{0}(\rho),\Lambda_{1}(\rho))
       \left(n_{0} \plus n_{1}\right)
       \left(\begin{array}{c}
            \! n_{0} \plus n_{1} \minus 2 \! \\
            \! n_{0} \minus 1 \!
       \end{array}\right)
   \equiv D\tilde{h}_{D}(\rho)B_{n_{0}n_{1}}
    \ \ ,                                               \label{eqn:curv_est}
\eqn
where $h_D$ captures the dependence on the seperation ratio $\rho$.
Fig.~3 shows the rescaled prefactors $\alpha_{n_{0}n_{1}}/DB_{n_{0}n_{1}}$ as
a function of $\rho$, together with the linear fits of the data:
\EqnA
   \tilde{h}_{1}(\rho)   \ApprA -0.584+0.378 \rho \\ [0.8ex]
   \tilde{h}_{1/2}(\rho) \ApprA -0.506+0.376 \rho
    \ \ .                                               \label{eqn:lin_fit}
\eqnA
The relative deviations from the linear fits are rather large for small
values of $\rho$, but shrink with increasing $\rho$; at $\rho=6$
the maximum relative error is less than 10\% (semiclassical case). The ansatz
$
   h_{D}(\Lambda_{0},\Lambda_{1})
   \approx\kappa_{D}\times|\Lambda_{0}\Lambda_{1}|^{1/2}
$
with $\kappa_{1}\approx\kappa_{1/2}\approx 0.2$ is another simple and
relatively accurate estimate.

\Eq{curv_est} gives us information with sufficient detail that we may now
return to the question whether there exist additional cancellations in
the classical and semiclassical cycle expansions for the open three disk
problem. The estimate of the full curvatures becomes
\EqnA
   \cjn{0}{n}
       \ApprA n D h_{D} \sum_{n_{0}=1}^{n-1}
       \left(\begin{array}{c}
               \!   n   \minus 2 \! \\
               \! n_{0} \minus 1 \!
       \end{array}\right)
       \lambda_{0}^{n_{0}}
       \lambda_{1}^{n-n_{0}}
       \EquAC   n D h_{D} \lambda_{0}\lambda_{1}
             \left(\lambda_{0}\plus\lambda_{1}\right)^{n-2}
          =     n D h_{D} \lambda_{0}\lambda_{1}f_{D}^{n-2}
       \ \ ,                                              \label{eqn:curv_est2}
\eqnA
where $f_{D}$ refers to $\fj{1}{2}$ in the classical and $\fj{1}{3}$
in the semiclassical case, respectively. For short, we have written
$
    \lambda_{0}\equiv t_{0}\Lambda_{0}^{-1}
$
and
$
    \lambda_{1}\equiv t_{1}\Lambda_{1}^{-1}
$.
The perhaps surprising observation to emerge from
\Eq{curv_est2} is that the leading pole
of $\zeta^{-1}_{0}(z)$ has to be double:
\Eqn
    \sum_{n} \cjn{0}{n}z^{n}
   = Dh_{D}\lambda_{0}\lambda_{1}f_{D}^{-1}z
    \sum_{n} n\left(f_{D}z\right)^{n-1}
   \Prop \frac{z}{\left(1-f_{D}z\right)^{2}} \ \ .  \label{eqn:double_pole}
\eqn
If now \EqRef{curv_est2} is inserted into the semiclassical
complexes \EqRef{canc_qm},
\Eqn
   \cjn{0}{n}-f_{D}\cjn{0}{n-1}
   \approx \left[n\minus\left(n\minus1\right)\right]
   Dh_{D}\lambda_{0}\lambda_{1}f_{D}^{n-2}
   = D h_{D}\lambda_{0}\lambda_{1}f_{D}^{n-2}
   \ne 0 \ \ ,                                                \label{eqn:rest}
\eqn
the terms do not cancel; there still remains a rest of (roughly)
the order $O(3n)$, building up a (simple) pole at
$
   z_{p}=1/\fj{1}{3}=
         \left(|\Lambda_{0}|^{-1/2}\Lambda_{0}^{-1}
         \plus|\Lambda_{1}|^{-1/2}\Lambda_{1}^{-1}\right)^{-1}
$.
This is in line with our earlier numerical findings that the
semiclassical Selberg zeta function is not free of poles.
However, if we insert the same expression \EqRef{curv_est2} into
the classical complexes \EqRef{canc_cl},
\Eqn
   \cjn{0}{n}-2f_{D}\cjn{0}{n-1}+f_{D}^{2}\cjn{0}{n-2}
   \approx
   \left[n \minus 2\left(n \minus 1\right)+\left(n \minus 2\right)\right]
   Dh_{D}\lambda_{0}\lambda_{1}f_{D}^{n-2}
   = 0 \ \ ,
\eqn
the additional cancellations are there.

Note that the qualitative results above are independent of the choice of weight
(i.e., the value of $D$); the different results for the classical and
the semiclassical Selberg zeta function are entirely due to the difference
in the power of $\zeta^{-1}_{1}$. The double pole of $\zeta^{-1}_{0}$ occurs at
$
  z_{p}\approx f_{D}^{-1}
$,
which is identical to the lowest order approximation of the leading zero of
$\zeta^{-1}_{1}(z)$. By taking the square of $\zeta^{-1}_{1}(z)$ as in the
classical Selberg zeta function one doubles the leading zero, which then
precisely cancels the double pole of $\zeta^{-1}_{0}$. In the semiclassical
case the simple zero of $\zeta^{-1}_{1}$ cancels only one pole with a simple
pole remaining [\Eq{rest}]. We study this point further below.

The conjecture \cite{AACI}
that the position of the poles of the dynamical
zeta function $\zeta_j$ is given by the zeros of $\zeta_{j+1}$
remains valid, but the order of the poles is not simple but
double. Returning to the arguments of Artuso {\em et al.}\/ \cite{AACI,Aurell}
one notes that they are rather liberal with the prefactors; and it is
precisely in the prefactors that the difference between a
simple and a double pole resides.

\section{Identification and removal of poles}

In the case of maps, phase space averages can be related to zeros of
zeta functions $Z(z)=\sum_{n=0}^{\infty}C_{n}z^{n}$.
Practical calculations estimate such zeros from a truncation of the
series, $F_{N}(z)=\sum_{n=0}^{N}C_{n}z^{n}$.
In ideal situations, exponential \cite{AACI,AACII}
or even faster than exponential convergence \cite{CCR}  is obtained.
The presence of poles destroys faster than exponential convergence and
makes it more difficult to calculate the exact positions of the
zeros; consider a simple case where there is one zero and one pole:
\EqnA
      F(z) \EquA \frac{1-az}{1-bz}
             =   \left(1-az\right)\left(1+bz+b^{2}z^{2}+\ldots\right)
           \EquAC 1-\left(a-b\right)z-\left(a-b\right)bz^{2}-\ldots
             = \sum_{n=0}^{\infty}C_{n}z^{n} \ \ .
                                                  \label{eqn:pole_ex}
\eqnA
We assume that $0<b<a$. A truncation after the linear term gives a value of
the zero $z'_{0}=\left(a-b\right)^{-1}$. This is obviously a bad estimate of the
true value $z_{0}=a^{-1}$ if $a$ and $b$ are of the same order of
magnitude. The inclusion of higher order terms only slowly
improves $z'_{0}$, the error being $\Prop(b/a)^{N}$ asymptotically.
Furthermore, additional ``ghost'' zeros appear: The number of these
unwanted zeros equals $N-1$ and they do not vanish to
infinity as $N$ grows large -- they cluster around the circle $|z|=b^{-1}$,
which borders the region of absolute convergence.
To see this, consider the function
\Eqn
   \tilde{F}_{N}(z)
    = { (1-az) (1-(bz)^{N}) \over 1-bz }
    = (1-az) \sum_{j=0}^{N-1} (bz)^j
    \ \ ,
                                                  \label{eqn:pole_ex_appr}
\eqn
which differs from $F_{N}(z)$
only in the coefficient $C_N$. It clearly has $N-1$ additional
zeros on the circle $|z|=b^{-1}$.

For $b>a>0$ the situation is still worse; there is {\em no}\/ zero
of $F_{N}(z)$ converging to $z_{0}=a^{-1}$.

If on the other hand the pole were absent, one would have a polynomial
$1-az$, which ``converges'' to its exact form already after the first
term in the ``expansion''; the zero $z_{0}=a^{-1}$ is at once correctly
determined. By estimating the value of $b$ from the asymptotic behavior
of the coefficients $C_{n}$, we could remove the effect of the
pole: Assume an estimate $\tilde{b}$ close to $b$ with
$|b-\tilde{b}|=|\varepsilon|\ll a$. Then the function
\Eqn
      (1-\tilde{b}z)F(z)
               = 1-\left(a-\varepsilon\right)z
                 -\varepsilon\left(a-b\right)z^{2}\left[
                 1+bz+b^2z^{2}+\ldots\right]
\eqn
has already in the linear approximation a zero
$\tilde{z}_{0}=\left(a-\varepsilon\right)^{-1}$ close to $z_{0}$.
The position of the ghost zeros may be estimated from the
function (cf.\ \Eq{pole_ex_appr})
\Eqn
     (1-\tilde{b} z)\tilde{F}_{N}(z)
   = (1-az) (1+(\varepsilon/b) \sum_{j=1}^{N} (bz)^j) \ \ ;
\eqn
for large z, it is dominated by the highest power of $z$, so that
its zeros lie on the circle
$
   |z| \Prop (\varepsilon/b)^{-1/N} b^{-1}
$.
They tend to infinity as $\varepsilon\rightarrow 0$.

Are several poles present, as e.g.\ in a rational function
\Eqn
       F(z)=\frac{P(z)}{Q(z)} \ \ ,
\eqn
with polynomials $P$ and $Q$, the zeros
$|z_{1}|\ll |z_{2}|\ll |z_{3}|\ll\ldots$ of $Q$ being different from the
zeros of $P$,
the removal of the leading pole $z_{1}$ leaves a function where $z_{2}$
determines the convergence. In the special case $P(z)=\left(1-az\right)$
and $Q(z)=\left(1-b_{1}z\right)\left(1-b_{2}z\right)$ with $0<b_{2}\ll b_{1}<a$,
the removal of $z_{1}=b_{1}^{-1}$ pushes the ghost zeros to the
neighborhood of $|z|=b_{2}^{-1}$ and the error in the linear estimate
of $z_{0}$ shrinks from $b_{1}\left[a\left(a-b_{1}\right)\right]^{-1}$,
which may be larger than $a^{-1}$, to
$b_{2}\left[a\left(a-b_{2}\right)\right]^{-1}\ll a^{-1}$.
If $z_{2}$ in the last example is not much larger than $z_{1}$, but
close in magnitude, the improvement is of course only marginal.
Thus, if two (or $n$) leading poles are close in magnitude, one has to
remove both (all $n$) before any considerable improvement takes place.

To remove a certain number of poles in a consistent way,
one may use a Pad\'{e} approximation \cite{Baker}:
Assume that $N+1$ coefficients $C_n$ of the function
$F(z)=P(z)/Q(z)=\sum_{n=0}^N C_n z^n$ are known and
that one wants to know the power series expansion of
the unknown functions $P_{M}(z)=\sum_{n=0}^{M}A_{n}z^{n}$
and $Q_{L}(z)=\sum_{n=0}^{L}B_{n}z^{n}$.
With the normalization $B_0=1$, one has to determine $M+1+L$
coefficients $A_n$, $B_n$, so that $N=M+L$ is required.
The coefficients are determined from $Q_{L}(z)F_{N}(z)=P_{N-L}(z)$, i.e.,
\Eqn
       \sum_{n=0}^{N}\sum_{n'=0}^{L}\theta(N \minus n \minus n')
          C_{n}B_{n'}z^{n+n'}
       = \sum_{n=0}^{N-L}A_{n}z^{n} \ \ ; \ \ \ \
       \theta(x)\equiv \left\{
      \begin{array}{ll} 0 & x<0 \\ 1 & x\ge 0 \end{array} \right.
      \ \ .
\eqn
By identifying terms of equal power in $z$, one finds altogether
$N+1$ equations for the $M+L+1=N+1$ unknowns.

Once the coefficients are known, the solution of $Q_{L}(\tilde{z}_{p})=0$
gives approximately the position(s) of the dominant pole(s).
If the computed value $\tilde{z}_{p}$ converges for
increasing $L$ (and $N$), one may feel confident about really having
identified a pole. Similarly, solutions of $P_{M}(\tilde{z}_{0})=0$ that
converge for increasing $L$ and $N$ should give improved estimates of the
zeros of the full function $F(z)$.
(Calculating poles is in general much more difficult than finding good
values for the zeros; even the leading pole requires large $N$.)

We have applied this to the three disk system. An approximation
with linear denominator $Q_{1}(z)$ already gave good results, confirmed and
stabilized by computations for quadratic and higher order denominators.
The positions of the leading zeros and poles of the full Selberg zeta
function and of the three first dynamical zeta functions are listed in
Tables~3-6. The data clearly confirm the conclusion drawn at the end of the
last section that the poles of $\zeta^{-1}_{j}\approx$ coincide with
zeros of $\zeta^{-1}_{j+1}$. The poles appeared in the computations either
as pairs of nearby real values or as pairs of complex conjugate values with
small imaginary parts; a strong indication that they are double.
Assuming that they are all real and double, the identification of poles
could in some cases be done with up to seven digits precision, applying an
extrapolation scheme on the results given by different orders of
approximations $N$ and $L$.

The improvement in the decay of the coefficients is apparent from
Fig.~4, where we show the the behaviour of the coefficients $A_{n}$
in the $P_{N-1}(z)/Q_{1}(z)$ approximation of the semiclassical
Selberg zeta function \EqRef{Zqm}; in line with the above discussion
we have approximated the denominator by $Q_1(z) = 1-z/z_1$, where
$z_1$ is the leading zero of $\zeta^{-1}_1$.
With the leading pole gone, the faster than exponential decay now continues
beyond $n=4$ out to $n=5$ or 6.

In the classical Selberg zeta function we were not able to identify any
pole with the Pad\'{e} technique ($N \le 13$), which would imply that they
are all cancelled by zeros of higher order dynamical zeta functions.
As our numerical results show, however, the (leading) poles of the dynamical
zeta functions are double; it is evident that the double zeros of
$\zeta^{-2}_{1}$ do cancel the double poles of $\zeta^{-1}_{0}$, as was also
shown in the preceding section, but the poles of
$\zeta^{-2}_{1}$ are then quadruple and too many to be completely
cancelled by the triple zeros of $\zeta^{-3}_{2}$. Thus we expect the classical
Selberg zeta function to have poles as well, the leading ones arising
not from $\zeta^{-1}_{0}$ as in the semiclassical case, but from
$\zeta^{-1}_{1}$. Since the magnitude of these poles is rather large, we
have not been able to extract them directly from a Pad{\'e} approximation
to $Z(z)$.

\section{Real and spurious zeros in quantum spectra}

As mentioned in the introduction most calculations require the zeta
functions not as functions of $z$ but rather as functions of
frequency $\omega$ and energy $E$ or wavenumber $k$.
Consider therefore a case similar to the one above, but with ``energy''
dependent coefficients, $C_{n}=C_{n}(E)$. Put $a=1$, $b=e^{-\delta}$,
and replace $z$ in \Eq{pole_ex} by $e^{iE}$ to obtain a function
\Eqn
      F(E) = \frac{1-e^{iE}}{1-e^{iE-\delta}} \ \ ;
\eqn
it mimics the behaviour of zeta functions in the case of
a bounded system.
We assume $\delta>0$, consistent with $0<b<a$ above. The function
$F(E)$ has zeros
$
   E_{n}=2\pi n$, $n\in N
$
along the real axis and poles
$
   E_{n'}=2\pi n'-i\delta$, $n'\in N
$
along the line $\mbox{Im}(E)=-\delta$, defining the abscissa of absolute
convergence in the complex $E$-plane.
We are interested in the properties of a finite order expansion of
$F$ and introduce therefore a generalized function $\hat{F}(E,z)$ with the
property $F(E)=\hat{F}(E,1)$,
\EqnA
      \hat{F}(E,z) \EquA \frac{1-e^{iE}z}{1-e^{iE-\delta}z}
                   \EquAC 1-(1\minus e^{-\delta})e^{iE}z
                     -(1\minus e^{-\delta})e^{-\delta}e^{2iE}z^{2}-\ldots
                   = \sum_{n=0}^{\infty}C_{n}(E)z^{n} \ .
                                                  \label{eqn:pole_ex2}
\eqnA
The expansion in $z$ is only formal; at the end one puts $z=1$.
All results of the former section concerning zeros, poles, etc.\ can now
be used, replacing $z$ by $e^{iE}$, $z_{0}$ by $e^{iE_{0}}$, etc.,
and relating them to the complex $E$-plane rather than the $z$-plane.
For a finite order expansion of $F(E)$ one may therefore immediately
state the following:

(i) Any truncation of the power series puts the zeros of $F(E)$ off
the real axis; in the linear approximation we obtain
$e^{iE}=\left(1-e^{-\delta}\right)^{-1}$, with solutions
$E'_{n}=E_{n}-i\epsilon$, where $\epsilon=-\log(1-e^{-\delta})$;
for $\delta$ sufficiently large $\epsilon\approx e^{-\delta}$.
The imaginary part of the energy vanishes with increasing $N$ like
$\epsilon\Prop e^{-\delta N}$. Large $N$ are required if the poles
lie close to the real axis.

(ii) Ghost zeros of $F(E)$ will be found in the neighborhood of the line
$\mbox{Im}(E)=-\delta$: There are $N-1$ of them
distributed more or less evenly around the circle $|e^{iE}|=e^{\delta}$;
they satisfy $e^{iE}=e^{i\phi_{j}+\delta_{j}}$ with
$\delta_{j}\approx\delta$, $0\le\phi_{j}<2\pi$, $j=1,2,\ldots,N-1$,
i.e., $E=E_{n,j}=E_{n}+\phi_{j}-i\delta_{j}$. The ghost zeros converge
towards the abscissa of absolute convergence and their number
goes to infinity with $N$.

One may remove poles in the same manner as before; the formal expansion
$\hat{F}(E,z)=\sum C_{n}(E)z^{n}$ is approximated with the Pad\'{e}
technique, $z$ is put to $1$, and left is a power series approximation
of the part of $F(E)$ containing the zeros.
The energy dependent analogue of the two-pole example in the former section
demonstrates the general tendency:

(iii) Removing leading poles has the following effect: Ghost zeros are
pushed down in the negative imaginary direction; main zeros having
small imaginary part approach the real axis.

The numerical investigations in this paper were performed for an
open hyperbolic system, and property (i) has no relevance.
Properties (ii)--(iii) are not restricted to bounded systems, though.
We present numerical evidence that the removal of the leading
pole does indeed push what seems to be ghost zeros in the open three
disk system far down in the negative complex energy plane: Fig.~5a shows the
lower part of the spectrum of the $N=8$ approximation of the semiclassical
Selberg zeta function together with an improved spectrum where the
leading pole has been removed (Pad\'{e}, $L=1$). The original spectrum displays,
in addition to the main zeros which lie relatively close to the real axis,
a whole band of lower lying zeros. When plotting spectra with higher $N$, one
obtains main zeros at the positions of the old ones, but the zeros in the band
seem not to stabilize and their number increases with $N$. In the improved
spectrum in Fig.~5a a few zeros in the band remain with approximately unchanged
or even larger imaginary part, while most of the others are pushed away;
in a certain range around $\mbox{Re}(k)\approx 13$, however, no such
improvement of the spectrum can be noted.

To find out whether the remaining zeros correspond to real resonances,
we can compare Pad\'{e} improved spectra
for different $N$. Fig.~5b shows improved ($L=1$) spectra for $N=6$
and $N=8$. Except for the zeros in the interval $\mbox{Re}(k)\approx 10-16$
most remaining zeros in the $N=8$ approximation correspond also to a nearby zero
in the $N=6$ approximation and thus seem to be stable. One exception is
the zero at $k\approx 24.4-i 1.5$ -- we have checked with the $N=10$
approximation though and have found a corresponding zero very close
to that position.

To see the correlation between the quantum spectra in Fig.~5a and the
position of the leading poles, we have plotted the absolute values of the
two lowest zeros of $\zeta^{-1}_{1}(E,z)$ [assuming that they
equal the poles of $\zeta^{-1}_{0}(E,z)$ and $Z(E,z)$] as a function of
$\mbox{Re}(k)$ with $\mbox{Im}(k)=-2.5$ fixed, see Fig.~6.
At low values of $\mbox{Re}(k)$ the magnitude of the second pole is much
larger than the leading one, so that the removal of the leading pole has a
large effect on the spectrum, which is also observed in Fig.~5a.
Around $\mbox{Re}(k)\approx 13$ the two poles are very close in magnitude,
which explains why the removal of just one pole only marginally affects
the spectrum. For increasing $\mbox{Re}(k)$ the number of leading poles
with comparable magnitude increases; it becomes more and more difficult
to improve the spectrum.

\section{Summary and discussion}

We have presented numerical evidence that the classical and semiclassical
Selberg zeta functions for the open three disk system have poles.
This shows that results for 1-d maps \cite{CCR,Pollicott}
cannot be transferred immediately to higher dimensional systems, such as
the three disk system. In the classical case
at least this is somewhat surprising, since the system under consideration
is an almost ideal hyperbolic 2d system (complete binary symbolics, highly
unstable periodic orbits with $|\Lambda_{p}|\gg 1$, no intermittency).
In the semiclassical case it was clear already from a simple plot of the
expansion coefficients that the convergence soon settles for a simple
exponential decay. This was confirmed by\\
\begin{tabular}{lp{12.5cm}}
(1)& a numerical determination of zeros and poles of dynamical
     zeta functions, showing that the leading poles of $\zeta^{-1}_{j}$
     equal the leading zeros of $\zeta^{-1}_{j+1}$, but are double and
     therefore not completely cancelled; \\
(2)& a numerical estimate of the curvature corrections of $\zeta^{-1}_{0}$,
     showing that the leading pole is double; \\
(3)& an analysis of the explicit terms in the expansion of $Z(z)$, which
     with use of (2) showed that necessary cancellations between cross terms,
     in addition to those present within the curvatures, do not occur,
     causing a transition in the convergence rate at around $n=4$; \\
(4)& the fact that stable poles could be extracted from $Q_{L}(z)$, the
     denominator of a Pad\'{e} approximation,
     $Z(z)\approx P_{N-L}(z)/Q_{L}(z)$,
     and that the numerator $P_{N-L}(z)$ showed improved convergence.
\end{tabular} \\
A plot of the expansion coefficients of the classical Selberg zeta function
shows seemingly a faster than exponential decay.
This is not in contradiction to (2)--(4) above as explained
at the end of section 2.3:
The poles of $\zeta^{-1}_{0}$ are cancelled by the zeros of
$\zeta^{-1}_{1}$, but there are not enough zeros from
$\zeta^{-1}_2$ to cancel the poles in $\zeta^{-1}_1$.
Nevertheless, we have not been able to locate a pole in the
classical Selberg product by Pad{\'e} analysis and
the tests of Cvitanovi{\'c} and Rosenqvist \cite{Priv} are
also consistent with an $\exp\{-n^{3/2}\}$ scaling.
The presence or absence of a pole in the classical Selberg zeta
function thus remains an open question.

Since the zeros of dynamical zeta functions are relatively easy to compute,
one may also quite easily identify the poles, provided there is a 1:1
(here rather 2:1) correspondence between poles and zeros of neighbouring
dynamical zeta functions. Our numerical findings give evidence that this is
most probably the case for the {\em leading}\/ zeros and poles. With the
precision of our investigations, we were not able to identify stable next to
leading poles from $Q_{L}(z)=0$ more than in a few cases and then only with
1--2 digits precision; in these cases there did exist a next to leading zero
within the uncertainty of the pole.

There are several ways to remove poles. The Pad\'{e} numerator $P_{M}(z)$
directly gives an approximation of $Z(z)$ or $\zeta^{-1}_{0}(z)$. One may
also first compute leading pole estimates $\tilde{z}_{n}$ from zeros of
dynamical zeta functions and then multiply
$
   \prod_{n}\left(1-z/\tilde{z}_{n}\right)
$
into the expansion of $Z(z)$. Furthermore, the classical Selberg zeta function
can be used with semiclassical weights to improve the convergence for
semiclassical zeros;
the poles of $\zeta^{-1}_{0}$ will then be gone.
If, finally, there actually is a $m:1$ correspondence between the
poles of $\zeta^{-1}_{j}$ and the zeros of $\zeta^{-1}_{j+1}$ for
all $j$, the following construction,
\Eqn
     \tilde{Z}(z)\equiv\prod_{j}\left[\zeta^{-1}_{j}\right]^{m^{j}}(z)
                 = \mbox{exp}\left\{
                   -\sum_{p}\sum_{r}
                    \frac{z^{rn_{p}}}{r}
                    \frac{t_{p}^{r}}{1-m\Lambda_{p}^{-r}}
                   \right\} \ \ ,                      \label{eqn:nopoles}
\eqn
is free of poles and has zeros which equal the zeros of $\zeta^{-1}_{0}(z)$
and $Z(z)$.
A numerical test for the three disk system ($m=2$) shows that this
zeta function (classical or semiclassical weights) really has faster than
exponential convergence all the way out to the largest $n$ considered (=13).
\Eq{nopoles} and related forms require further investigations.

Recently, quantum spectra of bounded chaotic systems have been computed
from expansions of the semiclassical Selberg zeta function
\cite{DR2,TSBEW,SS2}.
Despite some success, the calculations were made difficult by
slow convergence, no clear indication that the zeros of the Selberg zeta
function approach the real axis as $N\rightarrow\infty$, missing quantum
levels, and presence of spurious zeros (i.e., zeros of $Z(E)$ not associated
with exact quantum eigenvalues).
The pole induced properties (i)--(ii) of the preceding section
are here recognizable. We therefore suggest a relation,
similar to the simple example above, between the position of poles
and the distance from the real axis of the zeros of $Z(E)$
[or $\zeta_{0}^{-1}(E)$] in bounded systems.
In the calculations for the anisotropic Kepler problem and the
closed three disk system, there is one obvious pole, connected
with an orbit not realized by the dynamics,
but for which heteroclinic orbits of arbitrary length exist.
This pole has been removed in the calculations reported in \cite{TSBEW}.
But the present investigation suggests that there are further poles,
not so simply identified. We suspect that they are at least
partially responsible for the bad convergence of the zeros of $Z(E)$
(in bounded systems there are many other sources of trouble, like
intermittency and stable islands). Spurious zeros present in
cycle expanded spectra could be (in many cases at least) nothing but
the ghost zeros connected to the poles. Improved spectra should therefore
be obtained by removing the leading poles, as done here
for the open three disk system.

\section{Acknowledgments}

This work has been performed with financial support
(to G.~Russberg) from the Alexander von Humboldt-Stiftung.

\newpage


\newcommand{\PR}[1]{{\em Phys.\ Rep.}\/ {\bf #1}}
\newcommand{\PRL}[1]{{\em Phys.\ Rev.\ Lett.}\/ {\bf #1}}
\newcommand{\PRA}[1]{{\em Phys.\ Rev.\ A}\/ {\bf #1}}
\newcommand{\PRD}[1]{{\em Phys.\ Rev.\ D}\/ {\bf #1}}
\newcommand{\JPA}[1]{{\em J.\ Phys.\ A}\/ {\bf #1}}
\newcommand{\JPB}[1]{{\em J.\ Phys.\ B}\/ {\bf #1}}
\newcommand{\JCP}[1]{{\em J.\ Chem.\ Phys.}\/ {\bf #1}}
\newcommand{\JPC}[1]{{\em J.\ Phys.\ Chem.}\/ {\bf #1}}
\newcommand{\JMP}[1]{{\em J.\ Math.\ Phys.}\/ {\bf #1}}
\newcommand{\AP}[1]{{\em Ann.\ Phys.}\/ {\bf #1}}
\newcommand{\PLB}[1]{{\em Phys.\ Lett.\ B}\/ {\bf #1}}
\newcommand{\PD}[1]{{\em Physica D}\/ {\bf #1}}
\newcommand{\NPB}[1]{{\em Nucl.\ Phys.\ B}\/ {\bf #1}}
\newcommand{\INCB}[1]{{\em Il Nuov.\ Cim.\ B}\/ {\bf #1}}
\newcommand{\JETP}[1]{{\em Sov.\ Phys.\ JETP}\/ {\bf #1}}
\newcommand{\JETPL}[1]{{\em JETP Lett.\ }\/ {\bf #1}}
\newcommand{\RMS}[1]{{\em Russ.\ Math.\ Surv.}\/ {\bf #1}}
\newcommand{\USSR}[1]{{\em Math.\ USSR.\ Sb.}\/ {\bf #1}}
\newcommand{\PST}[1]{{\em Phys.\ Scripta T}\/ {\bf #1}}
\newcommand{\CM}[1]{{\em Cont.\ Math.}\/ {\bf #1}}
\newcommand{\JMPA}[1]{{\em J.\ Math.\ Pure Appl.}\/ {\bf #1}}
\newcommand{\RMP}[1]{{\em Rev.\ Mod.\ Phys.}\/ {\bf #1}}

\newpage

\section*{\center Figure captions}

\figlist{

  \item Expansion coefficients $C_{n}$ of the classical Selberg zeta
        function \EqRef{Zcl} for different $d/R$ in the open three disk
        system. ($A_1$-representation.)

  \item Expansion coefficients $C_{n}$ of the semiclassical Selberg zeta
        function \EqRef{Zqm} for different $d/R$ in the open three disk
        system. ($A_1$-representation.)

  \item Rescaled semiclassical (boxes) and classical (crosses) curvature
        prefactors,
        $
          \alpha_{n_{0}n_{1}}/DB_{n_{0}n_{1}}
        $,
        for different $\rho$ ($=d/R$) in the
        $A_1$-representation.
        The straight lines are the corresponding linear aproximations
	(Eq.~\EqRef{lin_fit}):
        $\tilde{h}_{1/2}(\rho)$ in the semiclassical case (broken line), and
        $\tilde{h}_{1}(\rho)$ in the classical case (continuous line).

  \item Improved convergence is achieved in the semiclassical Selberg
	zeta function \EqRef{Zqm} after removal of the leading pole:
        The figure shows the expansion coefficients of
        $(1-z/z_1)Z(z)$ where $z_1$ is the leading zero of
        $\zeta^{-1}_1(z)$. The data are for the $A_1$-representation.

  \item Semiclassical resonances for the open three disk system
        for $d/R=3$, computed using Selberg's zeta function
        in the $A_2$-representation.
        The energy is expressed in terms of $k=k(E)\equiv\sqrt{2mE}/\hbar$.
        (a) Spectrum in the $N=8$ approximation, without removal (boxes) and
            with removal (crosses) of the leading pole.
        (b) Spectrum with the leading pole removed, in the $N=8$ (boxes)
            and $N=6$ (crosses) approximation.

  \item The magnitude of the leading and next to leading zero of
        the semiclassical
        $\zeta^{-1}_{1}(E,z)$ as a function of $\mbox{Re}(k)$ with
        $\mbox{Im}(k)=-2.5$. The zeros are assumed to equal
        the leading and next to leading pole of $Z(E,z)$.
        ($d/R=3$, $A_2$-representation.)

}

\newpage

\section*{\center Table captions}

\tablist{

  \item Semiclassical curvature prefactors $\alpha_{n_0,n_1}$
	of $\zeta^{-1}_{0}$ (eq.~\EqRef{zqm})
        for the open three disk system ($A_1$-representation).

  \item Classical curvature prefactors $\alpha_{n_0,n_1}$
        of $\zeta^{-1}_{0}$  (eq.~\EqRef{zcl})
        for the open three disk system ($A_1$-representation).

  \item Leading zeros and poles of the semiclassical Selberg zeta function
	\EqRef{Zqm} and of
        the three lowest order dynamical zeta functions \EqRef{zqm} in the
        $A_1$-representation, determined from different Pad\'{e}
	approximations of the respective functions ($N\le 13$).

  \item Zeros and poles of the semiclassical zeta functions in the
        $A_2$-representation. (Cf.\ Table~3.)

  \item Zeros and poles of the classical zeta functions
	in the $A_1$-representation. Ellipses in column two indicate
	that no pole could be determined from the Pad\'e approximation.
	(Cf.\ Table~3.)

  \item Zeros and poles of the classical zeta functions
	in the $A_2$-representation. Ellipses in column two indicate
	that no pole could be determined from the Pad\'e approximation.
	(Cf.\ Table~3.)

}


\end{document}